\documentstyle[twoside,fleqn,espcrc2,epsfig]{article}

\newcommand{\AmS}{{\protect\the\textfont2
 A\kern-.1667em\lower.5ex\hbox{M}\kern-.125emS}}

\newcommand{\GeV}{\,{\rm GeV}}

\newcommand{\mum}{\,\mu{\rm m}}

\newcommand{\pom}{{\rm I\! P}}
\newcommand{\reg}{{\rm I\! R}}
\newcommand{\sgsp}{\sigma_{\rm tot}^{\gamma^*p}}
\newcommand{\sgp}{\sigma_{\rm tot}^{\gamma p}}

\newcommand{\ft}{F_2}

\newcommand{\fl}{F_L}

\newcommand{\pb}{\,{\rm pb}^{-1}}

\title{Measurement of the Proton Structure Function $\ft$ and of the
Total Photon-Proton Cross Section $\sgsp$ at Very Low $Q^2$ and Very
Low $x$}

\author{Christoph Amelung\address{Physikalisches Institut der
Universit\"at Bonn, Nu{\ss}allee 12, D-53115 Bonn,
Germany}\thanks{Supported by a grant from the Bundesministerium f\"ur
Wissenschaft und Forschung in Germany.} (ZEUS Collaboration)}

\begin{document}

\begin{abstract}

\noindent The proton structure function $\ft$ has been measured in the
range $0.045\GeV^2<Q^2<0.65\GeV^2$ and $6\cdot 10^{-7}<x<1\cdot
10^{-3}$ using $3.9\pb$ of $ep\to eX$ reactions recorded with the ZEUS
detector in 1997. The analysis is based on data from the Beam Pipe
Calorimeter (BPC) and the Beam Pipe Tracker (BPT). Compared to our
previous analysis, the BPT permits improved background suppression and
better control of systematic uncertainties, allowing the extension of
the kinematic region of the measurement towards lower $Q^2$ as well as
higher and lower $y$. Significant improvements have also been achieved
in the simulation of the hadronic final state via a mixture of samples
of non-diffractive and diffractive Monte Carlo events, generated by
the programs DJANGO and RAPGAP.

\end{abstract}

\maketitle

\section{INTRODUCTION}

Since the first measurement from ZEUS \cite{ZEUSBPC} using 1995 data,
the proton structure function $\ft$ at low $Q^2$ has continued to
generate a lot of interest. Various groups, among them ZEUS
\cite{ZEUSPHENO}, used the data to study the transition between
photoproduction and deep inelastic scattering, others obtained
improved parameterizations of $\ft$ \cite{ALLM97,DL98}.

At this workshop, a new measurement of $\ft$ is presented in the range
$0.045\GeV^2<Q^2<0.65\GeV^2$ and $6\cdot 10^{-7}<x<1\cdot 10^{-3}$,
where $x$ denotes the Bjorken scaling variable; $-Q^2$ is the
four-momentum transfer squared. One also defines $y$, the relative
energy transfer to the proton in its rest frame, and $W$, the
photon-proton center-of-mass energy. The data were taken with special
detector components and triggers during six weeks in 1997, yielding an
integrated luminosity of $3.9\pb$. The new analysis covers a larger
kinematic region and has a higher precision than the previous one
\cite{ZEUSBPC}.

\section{ANALYSIS}

\subsection{Scattered positron reconstruction}

The scattered positron is reconstructed in the Beam Pipe Calorimeter
(BPC) and Beam Pipe Tracker (BPT) of the ZEUS detector. The BPC has
been installed in 1995 and was used for the previous measurement of
$\ft$ at low $Q^2$. In 1997, the BPT was installed in front of the BPC.

The BPC is a small calorimeter that detects positrons with a
scattering angle of 1--2${}^{\rm o}$ w.r.t.\ the positron beam
direction. Its energy resolution is $\sigma_E=0.17\cdot\sqrt{E}$. The
energy scale is known to $\pm 0.3\%$, the non-linearity is less than
$\pm 1\%$ at $4\GeV$. The shower position in the BPC is
reconstructed with a resolution of $500\mum$ at $27.5\GeV$.

The BPT consists of two silicon microstrip detectors. A track is
reconstructed as the straight line through two hits in the BPT,
providing the positron scattering angle, its impact point on the BPC
and the event vertex. The uncertainty of the absolute BPT position is
less than $\pm 200\mum$. The tracking efficiency is known to $\pm 1.5\%$.

\begin{figure}[!htb] 

\epsfig{figure=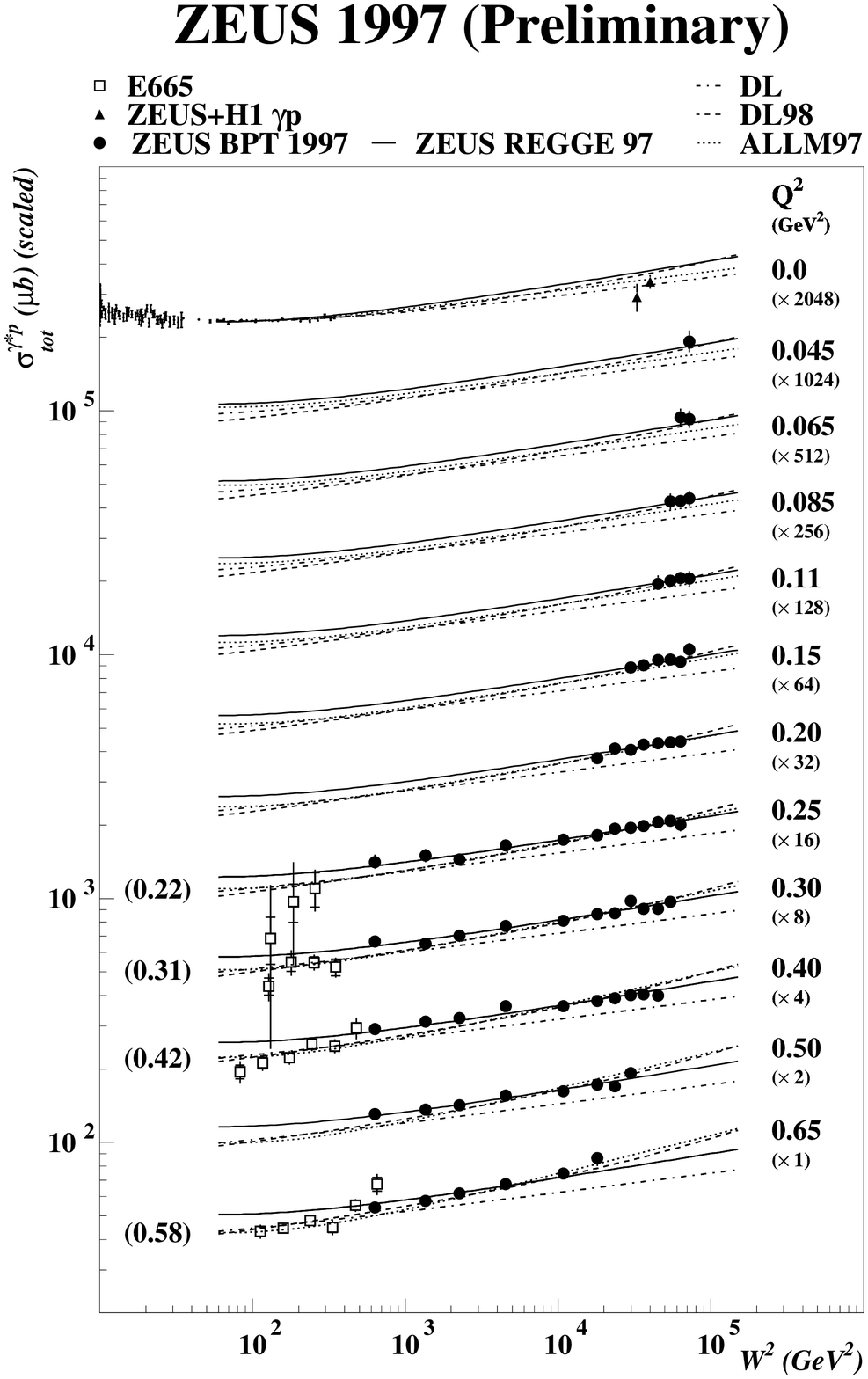,width=7.5cm}

\vspace{-0.8cm}

\caption{\label{SIGW}$\sgsp$ versus $W^2$ in bins of $Q^2$, compared
to E665 results and $\sgp$ from ZEUS and H1.}

\vspace{-0.5cm}

\end{figure}

\begin{figure}[!htb] 

\epsfig{figure=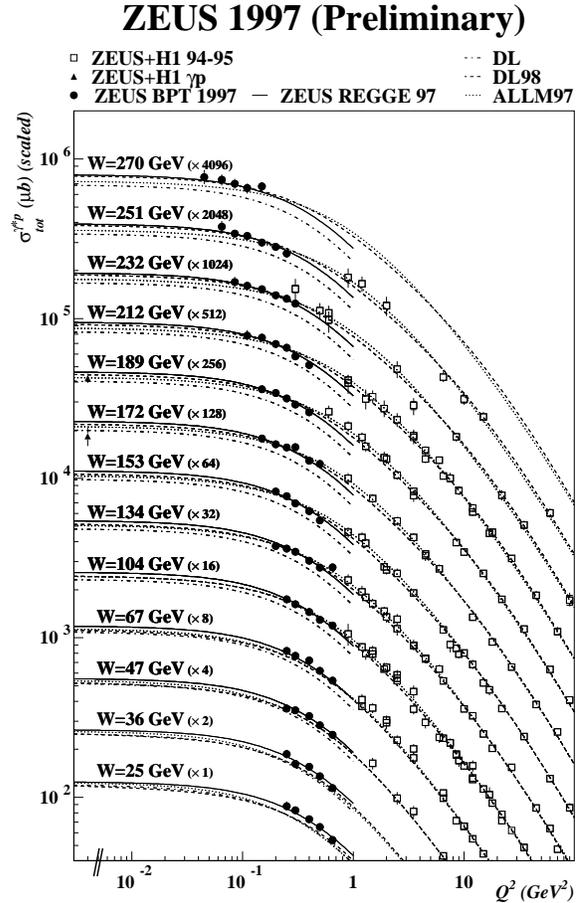,width=7.5cm}

\vspace{-0.8cm}

\caption{\label{SIGQ}$\sgsp$ versus $Q^2$ in bins of $W$, compared to
other measurements from ZEUS and H1.}

\vspace{-0.5cm}

\end{figure}

\subsection{Kinematic reconstruction}

The event kinematics are reconstructed with the electron method, i.e.\
from positron variables only, for $y>0.08$, where it gives the best
resolution. For $y<0.08$, the $e\Sigma$ method \cite{BASBERN} is used,
which improves the $y$ reconstruction by combining positron and hadronic final state
variables.

\subsection{Physics simulation}

DJANGOH 1.1 \cite{HSDJANGOH} and RAPGAP 2.06 \cite{HJRAPGAP} are used
to simulate non-diffractive respectively diffractive events. The
samples are mixed in a proportion determined from the data. This is
expected to give the best possible description of the hadronic final
state, which is crucial at high $y$ and low $Q^2$, where the fraction
of diffractive events rejected by trigger or offline cuts is
significantly different from non-diffractive events.

\subsection{Event selection}

The event selection is based predominantly on the requirement of a well
reconstructed positron in BPC and BPT. The analysis covers a kinematic
region between $0.045\GeV^2<Q^2<0.65\GeV^2$ and $6\cdot
10^{-7}<x<1\cdot 10^{-3}$, corresponding to $25\GeV<W<270\GeV$ or
$0.007<y<0.8$. Measuring at high $y$ was made possible by the use of the
BPT, which serves to suppress background at low scattered positron energies.

\section{RESULTS}

\subsection{\boldmath Determination of $\ft$ and $\sgsp$}

$\ft$ is extracted using an iterated bin-to-bin unfolding method and
converted into a total cross section via
$\sgsp={4\pi^2\alpha}/{Q^2}\cdot\ft$. The measured cross section
depends also weakly on $\fl$, which is taken from the BKS model
\cite{BKSMODEL}, yielding an effect of at most $3\%$. The results are
shown in figs.\ \ref{SIGW} and \ref{SIGQ}. In the region of this
analysis, the cross section becomes nearly flat as a function of
$Q^2$.

The data are displayed together with two fits (ALLM97 \cite{ALLM97},
DL98 \cite{DL98}) that have already included the 1995 measurement, as
well as with the older DL \cite{DL} curve. While DL98 gives the best
description at high $W$, it undershoots the data at low $W$. DL seems
to describe the shape better over the whole $W$ range. ALLM97
underestimates the cross section at low $Q^2$ by 10--15\%. The best
description of the data is given by the ZEUS REGGE 97 fit presented in
section \ref{PHENO}.

At low $W$, the analysis has reached kinematic overlap with
E665. While at $Q^2=0.65\GeV^2$ the agreement is good, it
deterioriates at lower $Q^2$.

\subsection{Systematic uncertainties}

Fifteen systematic checks were performed to study the stability of the
results. The average statistical error is $2.6\%$, the average
systematic error $3.3\%$. In most bins, the systematic error is
similar to the statistical one. Only the two highest $W$ bins are
dominated by systematic effects, mostly due to the uncertainty of the
fraction of diffractive events. The overall normalization uncertainty
of $\pm 1.8\%$ is due to the luminosity measurement.

\subsection{Phenomenological fits}
\label{PHENO}

With the precise $\ft$ data sample presented here, the
GVDM and Regge-inspired fits as published in \cite{ZEUSPHENO} have been
repeated.

The first step is to make an assumption about the $Q^2$ dependence of
the data in order to extrapolate them to $Q^2=0$, which is taken from the
GVDM prediction on $\sigma_T$, giving
$\sgsp(W^2,Q^2)={m_0^2}/{(m_0^2+Q^2)}\cdot\sgp(W^2)$. Fitting instead
both the $\sigma_T$ and $\sigma_L$ terms changes the extrapolated
values only within their statistical errors.

\begin{figure}[!htb] 

\epsfig{figure=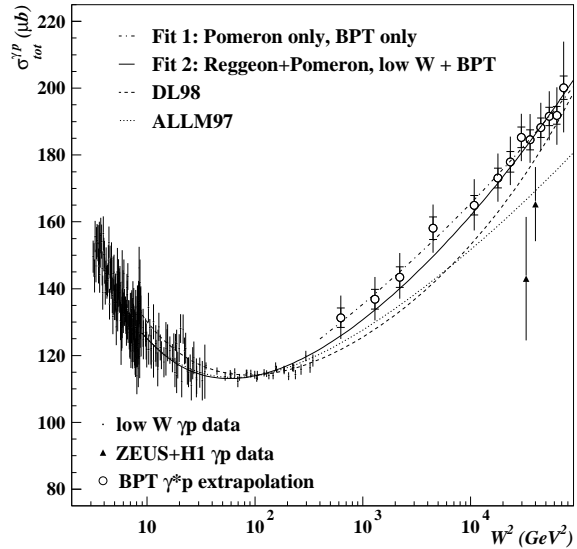,width=7.5cm}

\vspace{-0.9cm}

\caption{\label{FITW}Extrapolated $\sgp$ versus $W^2$, compared to the
two direct measurements from ZEUS and H1 and to data at lower energies.}

\vspace{-0.6cm}

\end{figure}

In the second step, the $W$ dependence of $\sgp(W^2)$ is explored. The
extrapolations from the first step are compared to the direct
photoproduction cross section measurements by ZEUS and H1
\cite{ZEUSSIGTOT,H1SIGTOT} and to previous data from other experiments
at lower $W$. This comparison is shown in fig.\ \ref{FITW}, together
with Regge-type fits of the form $\sgp(W^2)=A_\reg
W^{2(\alpha_\reg-1)}+A_\pom W^{2(\alpha_\pom-1)}$, and with the DL98
and ALLM97 parameterizations. The extrapolated cross sections are
larger than the directly measured ones. Whether this is a feature of
the assumed $Q^2$ dependence for the extrapolation or of the direct
cross section measurements, remains to be resolved in the future.

The combination of the two fits is shown as ZEUS REGGE 97 in figs.\ 
\ref{SIGW} and \ref{SIGQ}.

\section{CONCLUSIONS}

The ZEUS collaboration has measured the proton structure function
$\ft$ in the range $0.045\GeV^2<Q^2<0.65\GeV^2$ and $6\cdot
10^{-7}<x<1\cdot 10^{-3}$ with unprecedented precision. The data can
be described and extrapolated by a simple GVDM and Regge inspired
parameterization.

\end{document}